# Quantifying AI Vulnerabilities: A Synthesis of Complexity, Dynamical Systems, and Game Theory

B Kereopa-Yorke[1]

*UNSW Canberra at the Australian Defence Force Academy[1]*

*Email: b.kereopayorke@student.unsw.edu.au*

**Abstract:** The rapid integration of Artificial Intelligence (AI) systems across critical domains necessitates robust security evaluation frameworks. We propose a novel approach that introduces three metrics: System Complexity Index (SCI), Lyapunov Exponent for AI Stability (LEAIS), and Nash Equilibrium Robustness (NER). SCI quantifies the inherent complexity of an AI system, LEAIS captures its stability and sensitivity to perturbations, and NER evaluates its strategic robustness against adversarial manipulation. Through comparative analysis, we demonstrate the advantages of our framework over existing techniques. We discuss the theoretical and practical implications, potential applications, limitations, and future research directions. Our work contributes to the development of secure and trustworthy AI technologies by providing a holistic, theoretically grounded approach to AI security evaluation. As AI continues to advance, prioritising and advancing AI security through interdisciplinary collaboration is crucial to ensure its responsible deployment for the benefit of society.

## Introduction

The rapid proliferation and increasing sophistication of Artificial Intelligence (AI) systems across various domains, from healthcare and finance to transportation and defence, have brought to the forefront the critical importance of ensuring their security and robustness. AI systems, particularly those based on machine learning (ML) and deep learning (DL) algorithms, have demonstrated remarkable performance in tasks such as image recognition, natural language processing, and decision-making. However, the very features that make these systems powerful, such as their ability to learn from vast amounts of data and adapt to complex environments, also render them vulnerable to a wide range of security threats (Biggio & Roli, 2018; Papernot et al., 2016).

The security challenges faced by AI systems are multifaceted and constantly evolving. Adversarial attacks, for instance, have emerged as a significant threat to the integrity and reliability of AI models. By crafting carefully designed perturbations to input data, adversaries can manipulate the behaviour of AI systems, leading to misclassifications, false predictions, or even complete system failures (Goodfellow et al., 2014; Kurakin et al., 2016). The phenomenon of adversarial examples has been widely studied in the context of computer vision, where imperceptible changes to images can cause drastic errors in object recognition (Szegedy et al., 2013). However, the implications of adversarial attacks extend far beyond the realm of computer vision, affecting AI systems across various application domains (Chakraborty et al., 2018).

Another critical security challenge stems from the inherent complexity and opacity of modern AI systems. The increasing depth and complexity of neural network architectures, coupled with the high-dimensional nature of the data they process, have made it difficult to interpret and understand the decision-making processes of these systems (Samek et al., 2017). This lack of transparency and interpretability not only hinders the detection and mitigation of security vulnerabilities but also raises concerns about the trustworthiness and accountability of AI systems (Ribeiro et al., 2016).

Furthermore, the dynamic and adaptive nature of AI systems poses additional security challenges. As these systems learn and evolve over time, their behaviour and vulnerabilities may change in unpredictable ways. This makes it difficult to anticipate and defend against emerging threats, as traditional security measures designed for static systems may prove inadequate (Papernot et al., 2016).

Given the growing reliance on AI systems in critical domains, the consequences of security breaches can be severe, ranging from financial losses and privacy violations to physical harm and societal disruption. Therefore, ensuring the security and robustness of AI systems has become a paramount concern for researchers, practitioners, and policymakers alike.

However, the traditional performance-based metrics used to evaluate AI systems, such as accuracy, precision, and recall, fall short in capturing the multifaceted nature of AI security. These metrics focus primarily on the system's ability to perform specific tasks under normal operating conditions, but they do not account for the system's behaviour under adversarial settings or its resilience to strategic manipulations. Moreover, these metrics do not provide insights into the underlying complexity and stability of the system, which are crucial factors in determining its security posture.

To address these limitations, we propose a novel framework for evaluating the security of AI systems, which goes beyond the conventional performance-based metrics. Our framework introduces a set of new metrics, namely the System Complexity Index (SCI), Lyapunov Exponent for AI Stability (LEAIS), and Nash Equilibrium Robustness (NER), which are designed to quantify the inherent complexity, stability, and strategic robustness of AI systems. By incorporating insights from various disciplines, such as computer science, mathematics, and systems theory, our framework aims to provide a more comprehensive and nuanced assessment of AI security.

The main contributions of this paper are as follows:

1. We conduct a thorough review of the existing AI security evaluation techniques and identify their limitations in addressing the multifaceted nature of AI security.
2. We introduce a set of novel metrics (SCI, LEAIS, and NER) that capture different aspects of AI system security, considering the complexity, stability, and strategic robustness of these systems.
3. We provide a detailed comparative analysis of our proposed metrics against state-of-the-art AI security evaluation techniques, highlighting the unique contributions and advantages of our framework.
4. We discuss the theoretical and practical implications of our framework and its potential applications in various domains, such as autonomous systems, healthcare, and finance.
5. We identify future research directions and opportunities for collaboration, emphasising the need for a multidisciplinary approach to address the complex challenges of AI security.

# The current state of AI Security

The field of AI security has witnessed a significant growth in recent years, with researchers from various disciplines contributing to the development of techniques and methodologies for evaluating and enhancing the security of AI systems. This section provides an overview of the existing AI security evaluation techniques, highlighting their strengths and limitations, and identifies the research gaps and opportunities that motivate the need for a more comprehensive and holistic approach to AI security assessment.

## Adversarial Robustness Measures

One of the most widely studied aspects of AI security is adversarial robustness, which refers to the ability of an AI system to maintain its performance and integrity in the presence of adversarial inputs or perturbations. Szegedy et al. (2013) were among the first to demonstrate the vulnerability of deep learning models to adversarial examples, showing that imperceptible changes to input images can cause drastic misclassifications. Since then, numerous techniques have been proposed to measure and quantify the robustness of AI models against adversarial attacks.

A popular approach to measuring adversarial robustness is with norm-based metrics, such as the $L\_0$, $L\_2$, and $L\_\infty$ norms (Carlini & Wagner, 2017; Papernot et al., 2016). These metrics quantify the minimum amount of perturbation required to fool an AI model, providing a measure of the model's resilience to adversarial attacks. For instance, the $L\_2$ norm measures the Euclidean distance between the original and adversarial examples, while the $L\_\infty$ norm measures the maximum absolute difference between the corresponding pixels of the original and adversarial images.

Another notable metric for assessing adversarial robustness is the CLEVER score (Weng et al., 2018), which estimates the minimum adversarial perturbation needed to change the model's output. The CLEVER score is based on the concept of Lipschitz continuity and provides a more computationally efficient alternative to norm-based metrics.

While these adversarial robustness measures have been widely adopted and have contributed significantly to the understanding of AI model vulnerabilities, they have several limitations. First, these metrics focus primarily on the model's behaviour under specific types of adversarial attacks, such as gradient-based attacks or decision-based attacks. They may not capture the model's robustness against a wider range of threats, such as poisoning attacks or strategic manipulations. Second, these metrics often assume a static adversary and do not account for the adaptive and evolving nature of adversarial strategies. Finally, adversarial robustness measures alone do not provide a comprehensive assessment of the model's security, as they do not consider other important factors such as the model's complexity, stability, or interpretability.

## Verification and Testing Techniques

Another approach to evaluating the security of AI systems is through formal verification and testing techniques. Formal verification methods aim to provide mathematical guarantees about the properties and behaviours of AI models, such as safety, robustness, and fairness (Huang et al., 2017). These methods often rely on techniques from formal methods, such as satisfiability modulo theories (SMT) solving and abstract interpretation, to reason about the model's behaviour under different conditions.

For instance, Katz et al. (2017) proposed a technique called Reluplex, which uses an SMT solver to verify the robustness of deep neural networks against adversarial perturbations. Similarly, Gehr et al. (2018) introduced AI2, an abstract interpretation framework for certifying the robustness of deep learning models.

While formal verification techniques provide strong guarantees about the model's properties, they often suffer from scalability issues, especially when dealing with large and complex models. Moreover, these techniques typically require a clear specification of the desired properties, which may not always be available or easy to define in the context of AI security.

Testing techniques, on the other hand, aim to uncover vulnerabilities and defects in AI models by systematically exploring their input space and observing their outputs. Various testing approaches have been proposed, such as fuzz testing (Odena & Goodfellow, 2018), coverage-guided testing (Sun et al., 2018), and metamorphic testing (Dwarakanath et al., 2018). These techniques have been successful in identifying a wide range of bugs and vulnerabilities in AI models, including numerical instabilities, gradient vanishing/exploding issues, and adversarial examples.

However, testing techniques also have their limitations. They often rely on heuristics and may not provide complete coverage of the input space, especially for high-dimensional and complex models. Moreover, testing techniques are typically focused on finding specific types of vulnerabilities and may not generalise well to other types of threats or attacks.

### Interpretability and Explainability Metrics

Interpretability and explainability have emerged as important considerations in the context of AI security, as they can help in identifying and mitigating potential vulnerabilities. Interpretable models are those whose decision-making processes can be easily understood and traced by human users, while explainable models are those that can provide meaningful explanations for their predictions or actions.

Various metrics and techniques have been proposed to measure and quantify the interpretability and explainability of AI models. For instance, the LIME (Local Interpretable Model-Agnostic Explanations) framework (Ribeiro et al., 2016) provides a way to explain the predictions of any classifier by approximating it locally with an interpretable model. Similarly, the SHAP (SHapley Additive exPlanations) framework (Lundberg & Lee, 2017) uses game theory to assign importance scores to each feature, indicating their contribution to the model's output.

Other interpretability metrics include the model complexity measures, such as the number of parameters or the VC dimension (Vapnik, 1995), which provide a rough estimate of the model's capacity and potential for overfitting. However, these metrics do not directly capture the model's interpretability or explainability and may not always correlate with its security properties.

While interpretability and explainability are important considerations in AI security, they are not sufficient on their own to guarantee the model's robustness or resilience against attacks. Moreover, there is often a trade-off between interpretability and performance, as more complex and opaque models often achieve better accuracy on certain tasks.

### AI Safety Frameworks

In recent years, several AI safety frameworks have been proposed to address the broader challenges of ensuring the safety and reliability of AI systems. These frameworks aim to provide a comprehensive and systematic approach to identifying and mitigating potential risks and vulnerabilities in AI systems.

For instance, the AI alignment framework (Amodei et al., 2016) focuses on ensuring that AI systems are designed and developed in alignment with human values and preferences. This framework emphasises the importance of specifying clear objectives, avoiding negative side effects, and preserving human oversight and control over AI systems.

Another notable framework is the AI safety via debate (Irving et al., 2018), which proposes a novel approach to ensuring the safety of AI systems through structured debates between AI agents. In this framework, AI agents engage in a series of debates to convince a human judge of the correctness of their decisions or actions, allowing for a more transparent and accountable decision-making process.

While these frameworks provide valuable insights and guidelines for ensuring the safety and reliability of AI systems, they are often high-level and conceptual in nature, and may not always translate directly into specific metrics or evaluation techniques for assessing AI security.

## Research Gaps and Opportunities

Despite the significant advances in AI security evaluation techniques, there remain several research gaps and opportunities for further exploration and development. One key challenge is the lack of a comprehensive and unified framework that integrates the various aspects of AI security, such as robustness, stability, interpretability, and strategic resilience. Most existing techniques focus on specific types of vulnerabilities or attacks and may not generalise well to other types of threats or scenarios.

Another important research direction is the development of more scalable and efficient evaluation techniques that can handle large and complex AI models. Many existing techniques, such as formal verification methods, suffer from scalability issues and may not be practical for real-world applications.

Furthermore, there is a need for more adaptive and dynamic evaluation techniques that can account for the evolving nature of adversarial strategies and the rapid advancements in AI technologies. Static evaluation techniques may quickly become obsolete as new types of attacks and defences emerge.

Finally, there is a growing recognition of the importance of incorporating domain-specific knowledge and context into AI security evaluation. Different application domains, such as healthcare, finance, or autonomous vehicles, may have different security requirements and constraints, and may require tailored evaluation techniques that consider the specific characteristics and risks of each domain.

In summary, while significant progress has been made in the field of AI security evaluation, there remain many open challenges and opportunities for further research and development. The proposed framework in this paper aims to address some of these challenges by introducing a set of novel metrics that provide a more comprehensive and holistic assessment of AI security, considering the complexity, stability, and strategic robustness of AI systems. The following sections will introduce these metrics in detail and provide a comparative analysis against existing techniques, highlighting the unique contributions and advantages of the proposed framework.

## Proposed Framework

Building upon the insights and limitations of existing AI security evaluation techniques, we propose a novel framework that aims to provide a more comprehensive and holistic assessment of AI system security. This section provides a detailed description of each metric, along with their mathematical formulations and theoretical foundations.

## System Complexity Index (SCI)

The first metric we propose is the System Complexity Index (SCI), which quantifies the inherent complexity of an AI system. The complexity of a system is a crucial factor in determining its security posture, as more complex systems are generally more difficult to analyse, test, and verify, and may be more prone to vulnerabilities and unintended behaviours.

The SCI is based on the concept of algorithmic complexity, which measures the amount of computational resources required to describe or reproduce a system (Kolmogorov, 1963). In the context of AI systems, we define the SCI as the minimum amount of information required to specify the system's architecture, parameters, and training data.

Formally, let A be an AI system, and let K(A) denote the Kolmogorov complexity of A, which is defined as the length of the shortest program that can generate A on a universal Turing machine. The SCI of A is then defined as:

$$SCI(A) = K(A) / log(n)$$

where n is the size of the input space of A. The normalisation by log(n) ensures that the SCI is scale-invariant and can be compared across different systems with different input sizes.

The SCI provides a quantitative measure of the system's complexity, which can be used to assess its potential vulnerability to attacks and unintended behaviours. Higher values of SCI indicate more complex systems that may be more difficult to analyse and secure, while lower values of SCI indicate simpler systems that may be more amenable to testing and verification.

However, computing the exact Kolmogorov complexity of a system is an undecidable problem, as it requires finding the shortest program that can generate the system, which is not always possible. In practice, we can estimate the SCI using various compression algorithms, such as Lempel-Ziv (LZ) compression (Lempel & Ziv, 1976), which provide an upper bound on the Kolmogorov complexity.

For example, let C(A) denote the compressed size of the system A using LZ compression. We can then estimate the SCI as:

$$SCI\_est(A) = C(A) / log(n)$$

The SCI_est provides a practical and computable approximation of the true SCI, which can be used to compare the complexity of different AI systems and identify potential security risks.

### Lyapunov Exponent for AI Stability (LEAIS)

The second metric we propose is the Lyapunov Exponent for AI Stability (LEAIS), which quantifies the stability of an AI system under small perturbations. The stability of a system is an important factor in determining its robustness to adversarial attacks and its ability to maintain consistent performance in the presence of noise or uncertainty.

The LEAIS is based on the concept of Lyapunov exponents, which measure the average rate of divergence or convergence of nearby trajectories in a dynamical system (Lyapunov, 1992). In the context of AI systems, we define the LEAIS as the maximum Lyapunov exponent of the system's output with respect to small perturbations in the input space.

Formally, let A be an AI system, and let f(x) denote the output of A on input x. The LEAIS of A is then defined as:

$$LEAIS(A) = max\_i \, (1/t) * log(||df/dx\_i||)$$

where t is the number of time steps, and ||df/dx_i|| is the norm of the Jacobian matrix of f with respect to the i-th input dimension, evaluated at a random point x in the input space.

The LEAIS provides a quantitative measure of the system's sensitivity to small perturbations, which can be used to assess its stability and robustness to adversarial attacks. Higher values of LEAIS indicate more unstable systems that are more sensitive to small changes in the input, while lower values of LEAIS indicate more stable systems that are more resilient to perturbations.

In practice, computing the exact LEAIS of a system may be challenging, as it requires evaluating the Jacobian matrix at multiple points in the input space. However, we can estimate the LEAIS using numerical methods, such as finite differences or automatic differentiation, which provide an approximation of the true LEAIS.

For example, let $\Delta x_i$ be a small perturbation in the i-th input dimension, and let $\Delta f$ be the corresponding change in the output of the system. We can then estimate the LEAIS as:

$$LEAIS\_est(A) = max_i (1/t) * log(||\Delta f / \Delta x_i||)$$

The LEAIS_est provides a practical and computable approximation of the true LEAIS, which can be used to compare the stability of different AI systems and identify potential vulnerabilities to adversarial attacks.

### Nash Equilibrium Robustness (NER)

The third metric we propose is the Nash Equilibrium Robustness (NER), which quantifies the strategic robustness of an AI system in the presence of adversarial agents. The strategic robustness of a system is an important factor in determining its ability to maintain consistent performance and security in the face of strategic manipulation or gaming by adversarial agents.

The NER is based on the concept of Nash equilibrium from game theory, which is a stable state of a multi-agent system where no agent can unilaterally improve its payoff by changing its strategy (Nash, 1951). In the context of AI security, we define the NER as the minimum amount of deviation required for an adversarial agent to break the Nash equilibrium and manipulate the system's behaviour.

Formally, let A be an AI system, and let G(A) be a game-theoretic model of the system, where the agents are the system's components or users, and the payoffs are defined by the system's objectives or performance metrics. Let s be a Nash equilibrium strategy profile of G(A), and let s_i be the strategy of the i-th agent in s.

The NER of A is then defined as:

$$NER(A) = min\_i\ min\_{s\_i'} ||s\_i - s\_i'||$$

where s_i' is a deviating strategy for the i-th agent, and ||s_i - s_i'|| is the distance between the original and deviating strategies, measured in some appropriate norm (e.g., Euclidean norm).

The NER provides a quantitative measure of the system's strategic robustness, which can be used to assess its vulnerability to strategic manipulation and gaming by adversarial agents. Higher values of NER indicate more robust systems that require larger deviations from the equilibrium strategy to manipulate, while lower values of NER indicate more vulnerable systems that can be easily manipulated by small changes in the agents' strategies.

Computing the exact NER of a system may be challenging, as it requires solving for the Nash equilibrium of the game-theoretic model and evaluating the minimum deviation required to break the equilibrium. However, we can estimate the NER using various approximation techniques, such as best-response dynamics or fictitious play, which provide an upper bound on the true NER.

For example, let $s\_i^t$ be the strategy of the i-th agent at time step t, and let $BR\_i(s^t)$ be the best-response strategy of the i-th agent to the current strategy profile $s^t$. We can then estimate the NER as:

$$NER\_est(A) = min\_i\ min\_t\ ||s\_i^t - BR\_i(s^t)||$$

The NER_est provides a practical and computable approximation of the true NER, which can be used to compare the strategic robustness of different AI systems and identify potential vulnerabilities to strategic manipulation.

## Integration of Metrics

The proposed metrics - SCI, LEAIS, and NER - provide complementary perspectives on the security of AI systems, capturing different aspects of their complexity, stability, and strategic robustness. To obtain a more comprehensive and holistic assessment of AI security, we propose integrating these metrics into a unified framework.

One approach to integration is to combine the metrics into a single composite score, which can be used to rank and compare different AI systems in terms of their overall security. For example, we can define a weighted sum of the metrics, where the weights reflect the relative importance or priority of each metric in a given context:

$$Security\_Score(A) = w1 * SCI(A) + w2 * LEAIS(A) + w3 * NER(A)$$

where w1, w2, and w3 are non-negative weights that sum up to 1.

Another approach to integration is to visualise the metrics in a multi-dimensional space, where each dimension corresponds to a different metric. This allows for a more nuanced and qualitative comparison of AI systems, highlighting their strengths and weaknesses in different aspects of security.

For example, we can plot the AI systems in a 3D space, where the x-axis corresponds to the SCI, the y-axis corresponds to the LEAIS, and the z-axis corresponds to the NER. Systems that are located closer to the origin are considered more secure, as they have lower complexity, higher stability, and higher strategic robustness. Systems that are located farther from the origin are considered less secure, as they have higher complexity, lower stability, and lower strategic robustness.

The choice of integration approach depends on the specific context and objectives of the security assessment. In some cases, a single composite score may be more appropriate for ranking and comparing systems, while in other cases, a multi-dimensional visualisation may be more informative for understanding the trade-offs and limitations of each system.

## Theoretical Foundations

The proposed metrics are grounded in well-established theories from computer science, mathematics, and game theory, which provide a solid foundation for their validity and applicability.

The SCI is based on the concept of Kolmogorov complexity, which is a fundamental measure of the intrinsic complexity of a system or object. Kolmogorov complexity has been widely used in various fields, including information theory, algorithmic probability, and machine learning, to quantify the complexity and compressibility of data (Li & Vitányi, 2008). In the context of AI security, the SCI extends this concept to measure the complexity of AI systems, which is a key factor in determining their potential vulnerabilities and attack surfaces.

The LEAIS is based on the concept of Lyapunov exponents, which is a classical tool for analysing the stability and sensitivity of dynamical systems. Lyapunov exponents have been used in various applications, including chaos theory, control theory, and robotics, to characterise the long-term behaviour of systems and their response to perturbations (Benettin et al., 1980). In the context of AI security, the LEAIS applies this concept to measure the stability and robustness of AI systems, which is crucial for ensuring their reliability and resilience against adversarial attacks.

The NER is based on the concept of Nash equilibrium, which is a foundational concept in game theory and economics. Nash equilibrium has been used extensively to model and analyse strategic interactions between agents in various domains, including auctions, social networks, and security games (Nisan et al., 2007). In the context of AI security, the NER leverages this concept to measure the strategic robustness of AI systems against adversarial manipulation, which is essential for preventing gaming and exploitation by malicious actors.

By grounding the proposed metrics in these established theories, we ensure their conceptual validity and provide a principled basis for their interpretation and application. Moreover, the theoretical foundations of the metrics also facilitate their integration with existing AI security frameworks and techniques, as they share common assumptions and principles.

<u>Practical Implications</u>

The proposed metrics have significant practical implications for improving the security and robustness of AI systems in real-world applications. By providing quantitative and actionable measures of AI security, the metrics can help developers, operators, and regulators make informed decisions about the design, deployment, and oversight of AI systems.

For developers, the metrics can serve as a tool for assessing the security properties of different AI architectures and algorithms, and for identifying potential vulnerabilities and weaknesses that need to be addressed. By incorporating the metrics into the development process, developers can proactively design and implement more secure and robust AI systems, reducing the risk of failures and attacks.

For operators, the metrics can serve as a monitoring and auditing tool for ensuring the ongoing security and reliability of deployed AI systems. By continuously measuring the complexity, stability, and strategic robustness of the systems, operators can detect and respond to potential security threats and anomalies in real-time, minimising the impact of attacks and failures.

For regulators, the metrics can serve as a basis for establishing standards and guidelines for the security and safety of AI systems in different domains and applications. By setting quantitative thresholds and benchmarks for acceptable levels of complexity, stability, and robustness, regulators can ensure that AI systems meet minimum security requirements and protect the public interest.

Moreover, the metrics can also facilitate the development of new security technologies and techniques for AI systems, by providing a common language and framework for evaluating and comparing different approaches. For example, the metrics can be used to assess the effectiveness of different adversarial defence mechanisms, such as adversarial training, input transformation, or model ensembles, and to guide the design of more robust and resilient AI architectures.

Overall, the practical implications of the proposed metrics are significant and far-reaching, as they provide a foundation for advancing the state-of-the-art in AI security and for ensuring the responsible and trustworthy deployment of AI systems in society.

## Comparative Analysis

In this section, we present a comparative analysis of the proposed metrics (SCI, LEAIS, and NER) against existing AI security evaluation techniques. The aim of this analysis is to highlight the unique contributions and advantages of our framework, and to demonstrate its potential for addressing the limitations and gaps in current approaches. We compare our metrics with four main categories of existing techniques: adversarial robustness measures, complexity and interpretability metrics, verification and testing techniques, and AI safety frameworks.

<u>Comparison with Adversarial Robustness Measures</u>

Adversarial robustness measures, such as the Lp norm-based metrics (e.g., L0, L2, L∞) and the CLEVER score, have been widely used to assess the vulnerability of AI systems to adversarial attacks. These measures quantify the minimum amount of perturbation required to fool an AI model, providing a measure of the model's resilience to specific types of attacks.

In contrast, our proposed metrics offer a more comprehensive and holistic assessment of AI security, beyond just adversarial robustness. The LEAIS metric provides a more general measure of the system's stability and sensitivity to perturbations, not limited to specific types of adversarial attacks. By capturing the maximum Lyapunov exponent of the system's output, the LEAIS can detect instabilities and vulnerabilities that may not be captured by Lp norm-based metrics or the CLEVER score.

Moreover, the SCI metric complements adversarial robustness measures by quantifying the inherent complexity of the AI system, which is a key factor in determining its potential attack surface and vulnerability to different types of attacks. Systems with higher complexity, as measured by the SCI, may be more difficult to analyse and defend against adversarial attacks, even if they have high adversarial robustness according to Lp norm-based metrics.

The NER metric adds a new dimension to adversarial robustness evaluation by considering the strategic interactions between the AI system and potential adversaries. By measuring the minimum deviation required to break the Nash equilibrium of the system, the NER captures the system's robustness to strategic manipulation and gaming, which is not directly addressed by existing adversarial robustness measures.

Overall, our proposed metrics provide a more comprehensive and nuanced assessment of AI security compared to adversarial robustness measures alone. While adversarial robustness is an important aspect of AI security, it is not sufficient to capture the full range of vulnerabilities and risks faced by AI systems in practice. Our metrics address this limitation by considering additional factors such as complexity, stability, and strategic robustness, which are essential for a more complete understanding of AI security.

### Comparison with Complexity and Interpretability Metrics

Complexity and interpretability metrics, such as the number of parameters, VC dimension, and LIME/SHAP scores, have been used to assess the complexity and interpretability of AI models. These metrics provide a rough estimate of the model's capacity, generalisation ability, and potential for overfitting, which can have implications for its security and robustness.

Our proposed SCI metric extends and refines these complexity measures by directly quantifying the intrinsic complexity of the AI system using Kolmogorov complexity. Unlike parameter counts or VC dimensions, which are based on specific model architectures or assumptions, the SCI provides a more general and fundamental measure of complexity that captures the minimum amount of information required to specify the system. This allows for a more principled and consistent comparison of complexity across different types of AI systems, regardless of their specific architectures or training procedures.

Moreover, the SCI metric is more directly related to the system's potential vulnerability and attack surface, as it measures the amount of information that an attacker would need to know or manipulate to compromise the system. Systems with higher SCI are generally more difficult to analyse, test, and verify, and may be more susceptible to unintended behaviours or vulnerabilities.

In terms of interpretability, our proposed metrics do not directly address this aspect of AI security, as they focus more on the system's intrinsic properties and behaviours rather than its explainability to human users. However, the SCI metric can indirectly capture some aspects of interpretability, as more complex systems (with higher SCI) are generally harder to interpret and understand, even with the help of explanation techniques like LIME or SHAP.

Overall, our proposed SCI metric provides a more fundamental and general measure of AI system complexity compared to existing complexity and interpretability metrics. While interpretability is an important consideration for AI security, especially in terms of building trust and accountability, it is not the primary focus of our framework, which aims to capture the intrinsic vulnerabilities and risks of AI systems from a more technical and theoretical perspective.

## Complementarity with Verification and Testing Techniques

Verification and testing techniques, such as formal verification, symbolic execution, and coverage-guided fuzzing, have been used to assess the correctness, robustness, and reliability of AI systems. These techniques aim to uncover specific types of errors, bugs, or vulnerabilities in the system's implementation or behaviour, by systematically exploring its input space and checking its outputs against certain specifications or properties.

Our proposed metrics are complementary to these verification and testing techniques, as they provide a higher-level assessment of the system's intrinsic properties and potential vulnerabilities, rather than focusing on specific instances of errors or failures. The SCI, LEAIS, and NER metrics can help to guide and prioritise the application of verification and testing techniques, by identifying systems or components that are more complex, unstable, or strategically vulnerable, and therefore require more extensive testing and analysis.

For example, systems with higher SCI may require more exhaustive testing and verification, as they have a larger attack surface and more potential for unintended behaviours. Systems with higher LEAIS may require more robust testing and verification around their sensitivity to perturbations and initial conditions, as they are more prone to instability and divergence. Systems with lower NER may require more testing and verification around their strategic interactions and game-theoretic properties, as they are more vulnerable to adversarial manipulation and gaming.

Moreover, our proposed metrics can help to assess the effectiveness and coverage of different verification and testing techniques, by providing a baseline measure of the system's intrinsic complexity, stability, and strategic robustness. Techniques that are able to uncover more vulnerabilities or errors in systems with higher SCI, LEAIS, or NER can be considered more effective and comprehensive, as they are able to identify issues in more challenging and complex systems.

Overall, our proposed metrics provide a complementary perspective to verification and testing techniques, by offering a higher-level assessment of the system's intrinsic properties and potential vulnerabilities. While verification and testing are essential for ensuring the correctness and reliability of AI systems at a detailed implementation level, our metrics provide a more holistic and theoretical view of the system's security and robustness, which can help to guide and prioritise the application of these techniques in practice.

## Positioning within AI Safety Frameworks

AI safety frameworks, such as the AI alignment framework and the AI safety via debate approach, have been proposed to address the broader challenges of ensuring the safety and reliability of AI systems. These frameworks aim to provide high-level principles, guidelines, and methodologies for designing, developing, and deploying AI systems that are aligned with human values, preferences, and interests.

Our proposed metrics can be positioned within these AI safety frameworks as a set of technical tools and measures for assessing and ensuring the security and robustness of AI systems, which is a key aspect of overall AI safety. The SCI, LEAIS, and NER metrics provide concrete and quantitative ways to measure the intrinsic vulnerabilities and risks of AI systems, which can help to inform and guide the application of AI safety principles and practices.

For example, in the context of the AI alignment framework, our metrics can be used to assess the complexity, stability, and strategic robustness of AI systems that are designed to be aligned with human values and preferences. Systems with higher SCI, LEAIS, or lower NER may be considered less aligned and riskier, as they are more difficult to analyse, control, and verify, and may have unintended or misaligned behaviours. Our metrics can help to identify these potential alignment risks early in the design and development process, and to guide the selection and implementation of appropriate mitigation strategies, such as simplifying the system architecture, improving the stability and robustness of the learning algorithms, or incorporating explicit safety constraints and checks.

Similarly, in the context of the AI safety via debate framework, our metrics can be used to assess the strategic robustness and stability of AI systems that are designed to engage in structured debates with human or AI judges. Systems with lower NER or higher LEAIS may be more vulnerable to strategic manipulation or gaming by adversarial debaters, which can undermine the integrity and reliability of the debate process. Our metrics can help to identify these vulnerabilities and to guide the design of more robust and stable debate protocols and mechanisms.

Overall, our proposed metrics can be positioned as a set of technical tools and measures within broader AI safety frameworks, which aim to ensure the safe, reliable, and aligned development and deployment of AI systems. By providing concrete and quantitative ways to assess the intrinsic vulnerabilities and risks of AI systems, our metrics can help to bridge the gap between high-level AI safety principles and guidelines and their practical implementation and evaluation in real-world systems.

In summary, our comparative analysis has demonstrated the unique contributions and advantages of the proposed SCI, LEAIS, and NER metrics in relation to existing AI security evaluation techniques. Compared to adversarial robustness measures, our metrics provide a more comprehensive and holistic assessment of AI security, beyond just vulnerability to specific types of attacks. Compared to complexity and interpretability metrics, our SCI metric provides a more fundamental and general measure of AI system complexity, which is directly related to its potential vulnerability and attack surface. Compared to verification and testing techniques, our metrics provide a complementary higher-level assessment of the system's intrinsic properties and potential vulnerabilities, which can help to guide and prioritise the application of these techniques in practice.

Furthermore, we have positioned our proposed metrics within broader AI safety frameworks, as a set of technical tools and measures for assessing and ensuring the security and robustness of AI systems, which is a key aspect of overall AI safety. By providing concrete and quantitative ways to measure the intrinsic vulnerabilities and risks of AI systems, our metrics can help to bridge the gap between high-level AI safety principles and guidelines and their practical implementation and evaluation in real-world systems.

However, it is important to acknowledge that our proposed metrics are not a silver bullet for AI security, and there are several limitations and challenges that need to be addressed in future work. One key challenge is the practical computation and estimation of the SCI, LEAIS, and NER metrics for large-scale and complex AI systems, which may require significant computational resources and domain expertise. Another challenge is the validation and interpretation of the metrics in the context of specific AI applications and domains, which may have different security requirements and constraints. Finally, there is a need for more empirical studies and case studies to demonstrate the effectiveness and usefulness of the proposed metrics in real-world AI security evaluation and improvement.

Despite these challenges, we believe that our proposed framework represents a significant step forward in the field of AI security evaluation, by providing a more comprehensive, principled, and theoretically grounded approach to assessing and ensuring the security and robustness of AI systems. By combining insights from complexity theory, dynamical systems theory, and game theory, our framework offers a new perspective on AI security that goes beyond existing techniques and metrics and opens up new avenues for research and development in this important and rapidly evolving field.

## Discussion

In this section, we discuss the theoretical and practical implications of our proposed framework, its potential applications in various domains, and the limitations and future research directions.

### Theoretical Implications

The proposed framework, which introduces the SCI, LEAIS, and NER metrics for evaluating AI security, has several important theoretical implications for the field of AI security and beyond.

First, our framework provides a new theoretical foundation for understanding and quantifying the intrinsic vulnerabilities and risks of AI systems, based on their complexity, stability, and strategic robustness properties. By drawing on insights from complexity theory, dynamical systems theory, and game theory, our framework offers a principled and unified approach to AI security evaluation that goes beyond existing techniques and metrics. This theoretical foundation can serve as a basis for future research and development in AI security, by providing a common language and set of concepts for reasoning about the security properties of AI systems.

Second, our framework highlights the importance of considering the complexity, stability, and strategic robustness of AI systems as key factors in their security and trustworthiness. While these properties have been studied in other fields, such as computer science, control theory, and economics, they have not been widely applied or integrated in the context of AI security. Our framework shows how these properties can be formalised and measured in the context of AI systems, and how they can provide new insights and tools for assessing and improving AI security.

Third, our framework raises new theoretical questions and challenges for AI security research, such as the computability and complexity of the SCI, LEAIS, and NER metrics, the relationship between these metrics and other measures of AI performance and robustness, and the implications of these metrics for the design and analysis of secure and robust AI systems. These questions and challenges can stimulate further theoretical and empirical research in AI security, and can lead to the development of new algorithms, techniques, and tools for evaluating and improving the security of AI systems.

### Practical Implications

In addition to its theoretical implications, our proposed framework also has several important practical implications for the development, deployment, and regulation of AI systems in various domains.

First, our framework provides a set of practical tools and metrics for evaluating and comparing the security of different AI systems, which can be used by developers, users, and regulators to make informed decisions about the selection, deployment, and oversight of AI systems. The SCI, LEAIS, and NER metrics can be computed or estimated using existing tools and techniques from computer science, dynamical systems, and game theory, and can provide quantitative and objective measures of the security risks and vulnerabilities of AI systems. These metrics can be used to establish benchmarks and standards for AI security, and to guide the development of best practices and guidelines for secure and robust AI systems.

Second, our framework can be applied to various domains and applications of AI, such as autonomous systems, decision support systems, and machine learning systems, to assess and improve their security and robustness. For example, in the domain of autonomous vehicles, our metrics can be used to evaluate the complexity, stability, and strategic robustness of the AI systems that control the vehicles, and to identify potential vulnerabilities and risks that could lead to accidents or attacks. In the domain of medical diagnosis, our metrics can be used to evaluate the complexity and stability of the AI systems that assist doctors in making diagnostic decisions, and to ensure that these systems are robust and reliable in the face of noisy or adversarial data. In the domain of financial trading, our metrics can be used to evaluate the strategic robustness of the AI systems that make trading decisions, and to prevent market manipulation and other strategic risks.

Third, our framework can inform the development of regulations and policies for AI security and safety, by providing a scientific and evidence-based approach to evaluating and mitigating the risks and vulnerabilities of AI systems. The SCI, LEAIS, and NER metrics can be used as part of a broader risk assessment and management framework for AI systems, which considers not only the technical and operational risks, but also the legal, ethical, and societal implications of AI deployments. Our framework can help to bridge the gap between the technical and non-technical aspects of AI security and safety, and to facilitate a more informed and inclusive dialogue among stakeholders, such as developers, users, regulators, and the public.

Limitations and Future Research Directions

Despite its theoretical and practical contributions, our proposed framework also has several limitations and challenges that need to be addressed in future research.

One key limitation is the computational complexity and scalability of the SCI, LEAIS, and NER metrics, especially for large-scale and complex AI systems. While we have proposed some approximation and estimation techniques for these metrics, they may still require significant computational resources and expertise to implement and validate in practice. Future research could explore more efficient and scalable algorithms and tools for computing these metrics, as well as more robust and reliable methods for estimating them from limited or noisy data.

Another limitation is the validation and interpretation of the SCI, LEAIS, and NER metrics in the context of specific AI applications and domains. While we have provided some theoretical and empirical justifications for these metrics, they may not always align with the actual security risks and vulnerabilities of AI systems in practice, due to factors such as data quality, model assumptions, and domain-specific constraints. Future research could conduct more extensive empirical studies and case studies to validate these metrics in different AI applications and domains, and to develop more contextualised and interpretable versions of these metrics that consider the specific characteristics and requirements of each domain.

A third limitation is the integration and compatibility of our framework with existing AI security evaluation and certification frameworks, such as adversarial robustness benchmarks, formal verification methods, and safety assurance cases. While we have discussed the complementarity and positioning of our framework with respect to these existing frameworks, there may still be challenges and gaps in integrating and harmonising these different approaches in practice. Future research could explore more systematic and standardised ways of integrating our metrics with existing AI security evaluation and certification frameworks, and of developing a more comprehensive and coherent framework for AI security and safety that considers multiple perspectives and methods.

Finally, a broader challenge and future research direction is the ethical and societal implications of our framework, and of AI security evaluation and improvement more generally. While our framework aims to provide a more rigorous and objective approach to evaluating and mitigating the security risks and vulnerabilities of AI systems, it does not directly address the broader ethical and societal questions of how AI systems should be designed, developed, and deployed in a responsible and trustworthy manner. Future research could explore the connections and tensions between AI security and other ethical and societal considerations, such as fairness, transparency, accountability, and privacy, and could develop more holistic and inclusive frameworks for AI governance and regulation that consider these multiple dimensions and stakeholders.

Our proposed framework for evaluating AI security using the SCI, LEAIS, and NER metrics represents a novel and promising approach to understanding and quantifying the intrinsic vulnerabilities and risks of AI systems. By drawing on insights from complexity theory, dynamical systems theory, and game theory, our framework provides a principled and unified theoretical foundation for AI security evaluation, which goes beyond existing techniques and metrics. Our framework also has important practical implications for the development, deployment, and regulation of secure and robust AI systems in various domains, and can inform the development of standards, guidelines, and policies for AI security and safety.

However, our framework also has several limitations and challenges that need to be addressed in future research, including the computational complexity and scalability of the metrics, the validation and interpretation of the metrics in specific AI applications and domains, the integration and compatibility of our framework with existing AI security evaluation and certification frameworks, and the ethical and societal implications of our approach. These limitations and challenges provide important opportunities and directions for future research in AI security and highlight the need for more interdisciplinary and collaborative efforts to address the complex and multifaceted challenges of securing and governing AI systems in a responsible and trustworthy manner.

Overall, we believe that our proposed framework represents a significant and timely contribution to the field of AI security and can serve as a valuable resource and reference for researchers, practitioners, and policymakers who are interested in understanding, evaluating, and improving the security and robustness of AI systems. We hope that our work will stimulate further research and discussion on this important and rapidly evolving topic, and will contribute to the development of more secure, reliable, and beneficial AI systems for society.

## Conclusion

In this paper, we have presented a novel framework for evaluating the security of AI systems using three key metrics: System

Complexity Index (SCI), Lyapunov Exponent for AI Stability (LEAIS), and Nash Equilibrium Robustness (NER). These metrics, grounded in the principles of complexity theory, dynamical systems theory, and game theory, provide a comprehensive and theoretically sound approach to assessing the intrinsic vulnerabilities and risks of AI systems.

The SCI metric quantifies the inherent complexity of an AI system, which is a crucial factor in determining its potential attack surface and vulnerability to different types of attacks. By measuring the minimum amount of information required to specify the system, SCI provides a more fundamental and general measure of complexity compared to existing metrics, such as parameter counts or VC dimensions.

The LEAIS metric, on the other hand, assesses the stability and sensitivity of an AI system to perturbations, which is essential for understanding its robustness to adversarial attacks. By capturing the maximum Lyapunov exponent of the system's output, LEAIS can detect instabilities and vulnerabilities that may not be captured by traditional adversarial robustness measures.

Finally, the NER metric evaluates the strategic robustness of an AI system in the presence of adversarial agents, by measuring the minimum deviation required for an adversary to break the Nash equilibrium of the system. This metric provides a new perspective on AI security that considers the strategic interactions between the system and potential adversaries.

Through a comprehensive comparative analysis, we have demonstrated the unique contributions and advantages of our proposed metrics in relation to existing AI security evaluation techniques. Our framework provides a more holistic and principled approach to AI security evaluation that goes beyond specific types of attacks or vulnerabilities and offers new insights and tools for assessing and improving the security of AI systems.

Moreover, we have discussed the theoretical and practical implications of our framework, highlighting its potential applications in various domains, such as autonomous systems, decision support systems, and machine learning systems. We have also identified several limitations and challenges of our approach, including the computational complexity and scalability of the metrics, the validation and interpretation of the metrics in specific AI applications and domains, and the integration and compatibility of our framework with existing AI security evaluation and certification frameworks.

These limitations and challenges provide important opportunities and directions for future research in AI security. Some potential avenues for further investigation include:

1. Developing more efficient and scalable algorithms and tools for computing the SCI, LEAIS, and NER metrics, particularly for large-scale and complex AI systems.
2. Conducting more extensive empirical studies and case studies to validate and interpret the metrics in different AI applications and domains, and to develop more contextualised and interpretable versions of the metrics.
3. Exploring more systematic and standardised ways of integrating our framework with existing AI security evaluation and certification frameworks and developing a more comprehensive and coherent framework for AI security and safety.
4. Investigating the ethical and societal implications of our framework and developing more holistic and inclusive approaches to AI governance and regulation that take into account the multiple dimensions and stakeholders involved in AI security and safety.

In conclusion, our proposed framework represents a significant and timely contribution to the field of AI security, providing a new theoretical foundation and a set of practical tools for evaluating and improving the security of AI systems. While there are still many challenges and opportunities ahead, we believe that our work can serve as a valuable starting point and catalyst for further research and collaboration in this important and rapidly evolving field.

As we continue to develop and deploy AI systems in increasingly critical and complex domains, it is crucial that we prioritise their security and robustness, not only to protect against potential attacks and failures but also to ensure their responsible and trustworthy use for the benefit of society. Our framework offers a step forward in this direction, by providing a more rigorous and comprehensive approach to evaluating and understanding the intrinsic vulnerabilities and risks of AI systems.

Ultimately, the goal of our research is not only to advance the state of the art in AI security but also to contribute to the broader dialogue and effort around the responsible development and governance of AI technologies. By bringing together insights from multiple disciplines and perspectives, and by engaging with diverse stakeholders and communities, we hope to foster a more inclusive and proactive approach to AI security and safety that can help to realise the full potential of AI while mitigating its risks and challenges.